\pdfoutput=1
\documentclass{article}[12pt]
\newtheorem{proposition}{Proposition}

\usepackage{graphicx}
\title{An efficient reverse-lookup table based strategy for solving the synonym and cache coherence problem in virtually indexed, virtually tagged caches.}
\author{Madhav P. Desai, Aniket Deshmukh\\ Department of Electrical Engineering\\ IIT Bombay\\ Mumbai, India}
\begin{document}

\maketitle
\begin{abstract}

Virtually indexed and virtually tagged (VIVT) caches are an 
attractive option for micro-processor level-1 caches, 
because of their fast response time and because
they are cheaper to implement than more complex caches such as virtually-indexed
physical-tagged (VIPT) caches.   The level-1 VIVT cache becomes even simpler to
construct if it is implemented as a direct-mapped cache (VIVT-DM cache). 
However, VIVT and VIVT-DM caches have some drawbacks.  When the number of sets in the
cache is larger than the smallest page size, there is a possibility of synonyms (two or more virtual
addresses mapped to the same physical address) existing in the cache.  Further,
maintenance of  cache coherence across multiple processors requires a physical
to virtual translation mechanism in the hardware.  
We describe a simple, efficient reverse lookup table based approach to 
address the synonym and the coherence problems in VIVT (both set associative and direct-mapped)
caches.   
In particular, the proposed scheme does not disturb the critical memory access paths 
in a typical micro-processor, and requires a low overhead for its implementation.
We have implemented and validated  the scheme in the AJIT 32-bit microprocessor core 
(an implementation of the SPARC-V8 ISA) and the implementation uses 
approximately 2\% of the gates and 5.3\% of the memory bits in the processor core.
\end{abstract}

\section{Introduction}

\begin{figure}
  \centering
  \includegraphics[width=12cm]{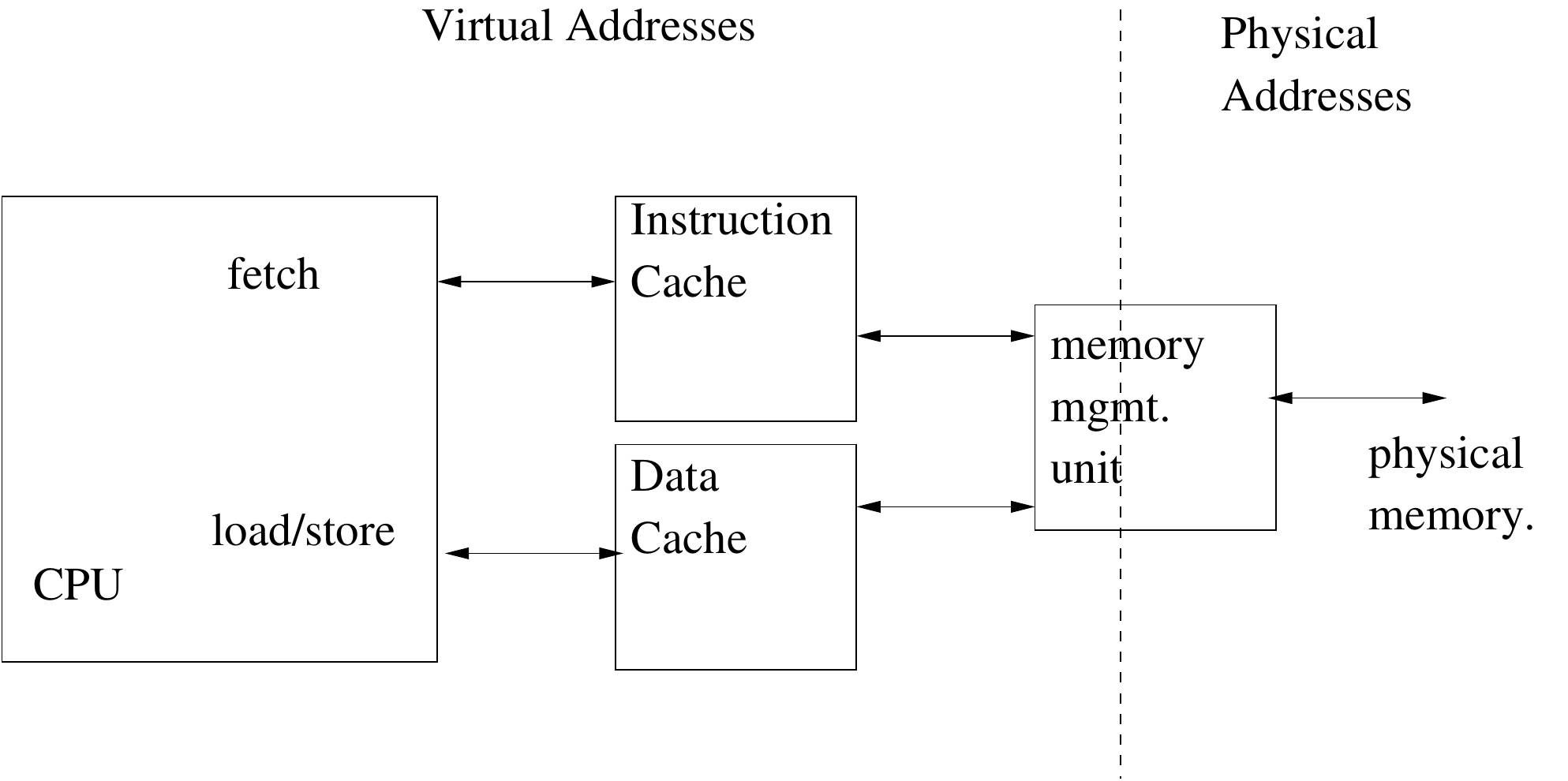}
  \caption{Processor with Virtually-indexed-virtually-tagged cache subsystem}
  \label{fig:TypVIVT}
\end{figure}

A typical processor with a virtually indexed, virtually tagged (VIVT) cache 
subsystem is shown in Figure \ref{fig:TypVIVT}.
The AJIT microprocessor developed in the Department of Electrical Engineering
at IIT-Bombay is an implementation of the SPARC-V8 (draft IEEE standard 1754)
instruction set architecture \cite{ref:SparcV8}, and this implementation uses
direct-mapped VIVT (VIVT-DM) caches.  This processor implementation will be used
as the concrete example of the use of the scheme outlined in this paper.

In VIVT caches, the cache lookup occurs before address translation, 
thus reducing the cache access latency on a hit, and also reducing
power dissipation by reducing the address translation activity in 
the memory management unit \cite{ref:VIVTSYNONYM0, ref:VIVTSYNONYM1}.  
In addition to a VIVT scheme, 
if the cache is organized in a direct mapped manner,   it is possible  to obtain 
single cycle cache read and write hit latencies by using single 
ported synchronous SRAM memory.   Even though
direct mapped caches can have higher miss ratios than set associative
caches, their speed of access on a hit and their low cost in terms of area and
power can be effective \cite{ref:DirectMappedCache}.
We will refer to VIVT direct-mapped caches as VIVT-DM caches.

The other popular cache organization which is commonly used in microprocessors
is the virtually-indexed, physically tagged (VIPT) cache.   In order to 
prevent aliasing (that is there must exist at most one entry for a physical
address) in a VIPT cache, it is necessary that two virtual addresses
which map to the same physical address must have the same set index.
This implies that the cache associativity must
be equal to the cache size divided by the page size.  Thus, with 4 Kilobyte pages,
a VIPT cache with 32 Kilobyte capacity will need to be 8-way set associative.  

VIVT caches also have drawbacks.  It is possible that
in different contexts, the same virtual address is mapped to
two different physical addresses.  This corresponds to a situation
where the virtual address is used by two distinct processes (contexts) with two
distinct page mappings.    This problem can be solved by
flushing the VIVT caches (both instruction and data) every time there is a context switch, or
by including context information in the cache tag \cite{ref:VIVTSYNONYM0}.

The second problem is that of synonyms.  Even in the same context, it
is possible to have two virtual addresses map to the
same physical address.   In such a situation, if the number of sets in 
the cache is larger than the minimum page size, it is possible that there are 
multiple virtual addresses in the cache which map to the same physical address.  
This can result in lack of consistency in the cache.  In the literature, proposed solutions
to the synonym problem fall under the following classes 
\cite{ref:VIVTSYNONYM0, ref:VIVTSYNONYM1, ref:UCACHE}:
\begin{itemize}
\item Software solutions: control of virtual to physical mappings, 
forcing synonyms to collide in the cache, using protection mechanisms
to allow read-only synonyms.
For example, the VIPS-M protocol \cite{ref:VIVTSYNONYM3}
relies on a classification of data at a page granularity as private or shared,
using the operating system and  and the translation look-aside buffer.   Management
of shared pages is done via interrupt service routines and manipulation of the translation lookaside
buffers in the cores.   
\item Hardware solutions: use a reverse lookup table to track 
synonyms in the cache \cite{ref:UCACHE}.  Typically such approaches
try to keep track of synonyms that exist in the cache and 
to maintain consistency in the cache by using this synonym information.
\end{itemize}

The third problem is the maintenance of cache coherency in
a multi-core system.    In a multi-core
scenario, a snoop-based cache coherency protocol will generate
invalidation messages identifying a physical address to be purged
from the cache \cite{ref:SnoopProtocols}.  In the VIVT-DM case, this physical address needs
to be translated to the (possibly multiple) virtual address(es)
which map to this physical address.   It is critical that this
reverse translation is achieved with low latency and 
high throughput so that the cache coherence related invalidation activity does not
itself become a bottleneck in the performance of the overall
system.

We provide a unified solution which uses a reverse lookup table
to tackle the problem of synonyms and coherency maintenance in an 
elegant and cost effective manner.  We fix an integer parameter 
$S \in \{1,2,\ldots\}$ and ensure that the following invariant is always maintained.
\begin{verbatim}
     For every physical address for which data is present 
     in the cache,  there are at most S virtual addresses
     in the cache which map to it. 
\end{verbatim}  
This invariant is the basis of the 
simplicity and efficiency of the scheme.   In our approach, we ensure
that there are never more than $S$ synonyms for a particular physical
address. This makes the maintenance of consistency in the cache easy, 
and also simplifies the application of snoop related invalidates considerably.
We call a VIVT cache with this property an {\em $S$ synonym safe cache}.

Note that there is a trade-off between the read hit ratio and invalidation overhead
implied by the choice of $S$.  A small value of $S$ makes it easy and fast
to apply invalidations triggered by the cache coherency mechanism or by writes, 
whereas a large value of $S$ increases the read hit ratio 
by keeping all synonyms active in the cache.   In our current implementation in
the AJIT micro-processor, we have maintained $S=1$ because 
synonyms are rare in most user software and the hit rate degradation due to
synonym eviction is not a serious enough issue to require maintaining values
of $S > 1$.   

In the rest of this paper, we will describe the scheme with
the AJIT processor implementation as a reference.    The general
scheme can be applied to set associative caches as well as to direct mapped
caches.  Concretely, we have applied and validated it on the 
VIVT-DM caches used in the AJIT processor.

\section{Conceptual basis}

    We assume throughout that homonyms are not possible either
    by explicit flushing of the data and instruction caches on context switches, 
    or by including the context as part  of the cache tag.  

    We observe the following.

\begin{proposition} \label{prop:setassoc}

     Consider a homonym free VIVT set associative cache of size 
     $32$ kilo-bytes, with a cache line size of $64$ bytes, a physical
     address width of $36$, and a virtual address width of $32$, and a set size $2^r$ with
     $ 0 \leq r < 3$ (for this cache configuration, if $r \geq 3$, the synonym problem does not arise at all).   
     Assume that the page size in the virtual to physical translation mechanism is 4 kilo-bytes.

     For any address, we denote bits [11:0] of that address as the intra-page offset of 
     the address.
     \begin{enumerate}
	\item For a given intra-page offset, there can be at most $8/{2^r}$
		physical addresses whose corresponding data is present in the VIVT set
		associative cache.
	\item For a physical address $P$ whose data is present in the VIVT set associative cache, 
		there can be at most $8/{2^r}$ virtual addresses which map to this $P$, and for every
		such virtual address $V$ we must have $V[11:0] = P[11:0]$.  Consequently,
		the set index in the VIVT set associative cache at which $V$ resides is 
		entirely determined by the concatenation of $V[(14-r):12]$ and $P[11:6]$.
    	\item  To maintain the physical to virtual reverse lookup in this VIVT cache, 
    		we can use a $8/{2^r}$-way set associative reverse lookup table, 
		with the set index being determined by the bits $[11:6]$ of the physical address, and 
		bits $[35:12]$ being used as the tag.   The set indices (in the VIVT set associative
		cache) of the virtual addresses which map 
    		to $P$ can be entirely determined by keeping bits $[(14-r):12]$ of the virtual addresses 
		as data corresponding to the entry for $P$ in the set associative memory.  
     \end{enumerate}
\end{proposition}
{\bf Proof:}
    For a cache entry
    corresponding to a 32-bit virtual address $V$, let $P$ be the 36-bit physical address to which it
    is mapped.   Then, since virtual pages are mapped to physical pages, the intra-page offset
    of the virtual address and the physical address to which it is mapped must be the same.  That is,
    \begin{displaymath}
    V[11:0] = P[11:0]
    \end{displaymath}
    In a 32 kilo-byte VIVT set associative cache with set size $2^r$ and with 64-byte cache lines, 
	there can be 512 lines present in 
    	the cache.  The index of the set in which the line is stored in the cache is calculated 
    	by using bits [(14-r):6] of the virtual address.  Consequently, for a given
        intra-page offset, there can be at most $8/{2^r}$ possible virtual addresses 
	(determined by bits [(14-r):12]) present in the cache with this intra-page offset.
    Thus, since there are no homonyms, there can be at most $8/{2^r}$ possible physical addresses
    with a given intra-page offset in the cache at any time.  This completes the proof of the first statement in Proposition \ref{prop:setassoc} 

    Now, if $P$ is a physical address to which some virtual address $W$ in the VIVT set associative cache is
    mapped, then we must have $W[11:0] = P[11:0]$.  The set index of $W$ is 
    $W[(14-r):6]$.   Since bits $W[11:0]$ are fixed, there are at most $8/{2^r}$ potential values
    of $W$ (determined by $W[(14-r):12]$) which can be present in the cache.    
    This proves the second statement in Proposition \ref{prop:setassoc}. 

    Finally, to maintain the physical to virtual reverse lookup in the VIVT set associative
    cache, we can use an $8/{2^r}$-way set associative reverse lookup table, with the set index being determined
    by the bits $[11:6]$ of the physical address, and bits $[35:12]$ being used as the tag.   
    The  set indices (in the VIVT set associative cache) of the virtual addresses which map 
    to $P$ are entirely determined by keeping bits $[(14-r):12]$ of the virtual addresses as data 
    corresponding to the entry for $P$ in the reverse lookup table.

\vspace{1cm}
{\bf Note:}  If the VIVT cache is direct mapped (that is $r=0$ in the discussion above), then
it is sufficient to keep just VA[14:12] as the data corresponding to a physical address
entry in the RLUT.   If on the other hand, $r > 0$, then we need to keep VA[31:12] as the
data corresponding to the physical address in order to resolve the virtual address in 
the VIVT set associative cache.  Thus, there is a substantial saving in RLUT memory if
a VIVT-DM cache is used in the processor.

\subsection{Maintenance of the invariant}

    We turn our attention to the maintenance of the invariant.
    Assume that at a particular point in time, the invariant holds.
    Suppose that a virtual address $V$ is presented to the VIVT cache,
    and this address is not present in the cache.   This triggers an address translation
    action in the memory management unit (MMU).   Suppose the physical address generated
    by the translation is $P$.  Now, based on our invariant, there can be at most
    $S$ virtual addresses present in the VIVT cache which map to $P$.  

    Let us say there
    are $S$ virtual addresses $W_1$, $W_2$, $\ldots$, $W_S$ present in the VIVT cache 
    and each that $W_i$ maps to $P$.   The MMU presents
    $P$ to the reverse lookup table (RLUT) and this returns the list $W_1, W_2, \ldots W_S$ 
    as the virtual addresses already present in the cache.   To preserve the invariant, we must
    \begin{enumerate}
    \item Replace one of the $P \rightarrow$ entries from the RLUT  by 
    	  	the reverse translation $P \rightarrow V$.
    \item Invalidate the VIVT line corresponding to the replace virtual address.
    \item Further, if this is a write miss, then every non-evicted virtual address among
		the $W_1, W_2, \ldots W_S$ is also invalidated in the VIVT cache.
    \end{enumerate}

    If on the other hand, the number of addresses returned by the RLUT is
    $T < S$, then we 
    \begin{enumerate}
    \item Add the translation $P \rightarrow V$ to the RLUT.
    \item If this is a write miss, we ask the VIVT cache to invalidate 
           all the virtual addresses which match $P$.
    \end{enumerate}

    The same reverse lookup table can be used to process invalidation messages
    generated by a snooping cache coherence protocol (which will present a 
    physical address to invalidate a cache entry).  The only difference is
    that if there is a match of $P$ to $W$ in the RLUT, then a message to
    invalidate $W$ is always sent to the VIVT cache.

    To summarize: 
    \begin{itemize}
    \item At most $S$ synonyms are maintained in the VIVT cache.  
    \item A reverse lookup table is used to translate a physical
  	   address to a virtual address resident in the cache in order to
	   invalidate it.  For every physical address presented to the RLUT,
           there can be at most one virtual address in the RLUT which matches
	   this physical addres.
    \end{itemize}

\section{Implementation of the scheme in the AJIT processor}

The AJIT processor uses VIVT-DM caches with a cache size of 32 kilo-bytes, a minimum
page size of 4 kilo-bytes and a
line size of 64 bytes.

A memory access generated in the CPU of the AJIT processor goes through
the sequence of actions shown in Figure \ref{fig:MemAccess}.
\begin{figure}
\begin{verbatim}
mem_access(rwbar, addr, wdata) {
     // access tags and data in parallel
     hit, rdata = hitCheckAndAccessCache(rwbar, addr,wdata);
     if (hit) {
         if (rwbar) {
             writeThroughToMem(addr,wdata);
             if (S > 1) {
                // Note: this slows down write-through
                // on a hit.  We keep S = 1 in the AJIT
                // processor.
                applyRlutInvalidates();
             }
         }
         return(rdata);
     }
     else
     {
         // includes write-through
         error, line = accessCacheLine(rwbar,addr,wdata);
         if(!error)  
            updateCache(addr, line);
     } 
}
\end{verbatim}
\caption {Memory access.}
\label{fig:MemAccess}
\end{figure}

As can be seen from Figure \ref{fig:MemAccess}, the AJIT processor
cache is direct-mapped and uses a write-through allocate policy.  
The tag and data arrays are updated together, thus minimizing
the access latency on a hit.  On a miss, a new line is fetched
from the main memory (after address translation in the MMU)
and the cache is left in the desired state, ready for the next
access.  Note that keeping $S=1$ is especially beneficial in the
write-through on hit scenario.

\begin{figure}
  \centering
  \includegraphics[width=12cm]{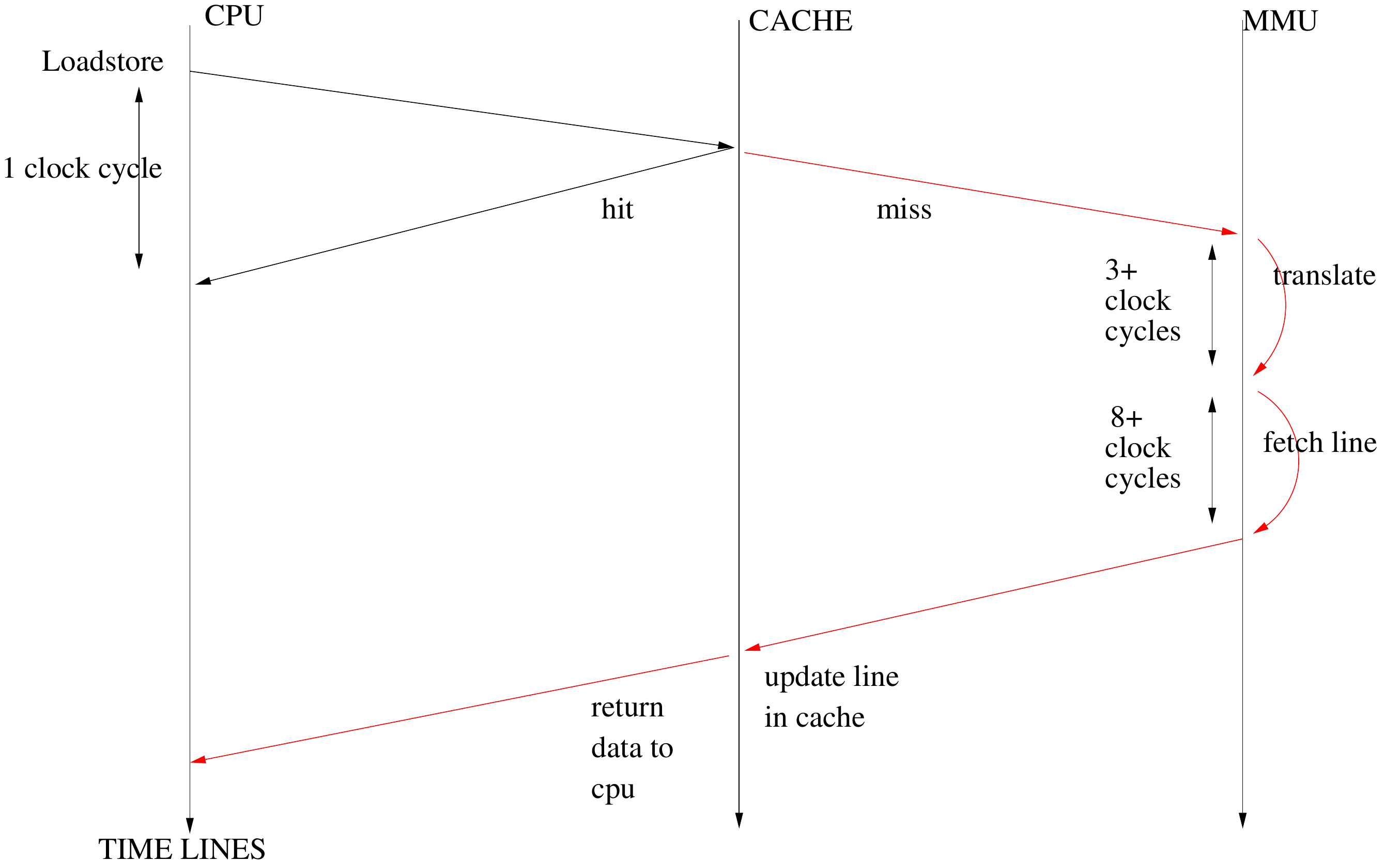}
  \caption{Activity in the memory access pipeline.}
  \label{fig:MemAccessFlow}
\end{figure}

A time-line view of this scheme is shown in Figure \ref{fig:MemAccessFlow}.
We observe that the on a miss, there are two sequential actions in the
MMU.  First an address translation is performed, and the translated address
is used to fetch a new cache line.

\begin{figure}
  \centering
  \includegraphics[width=12cm]{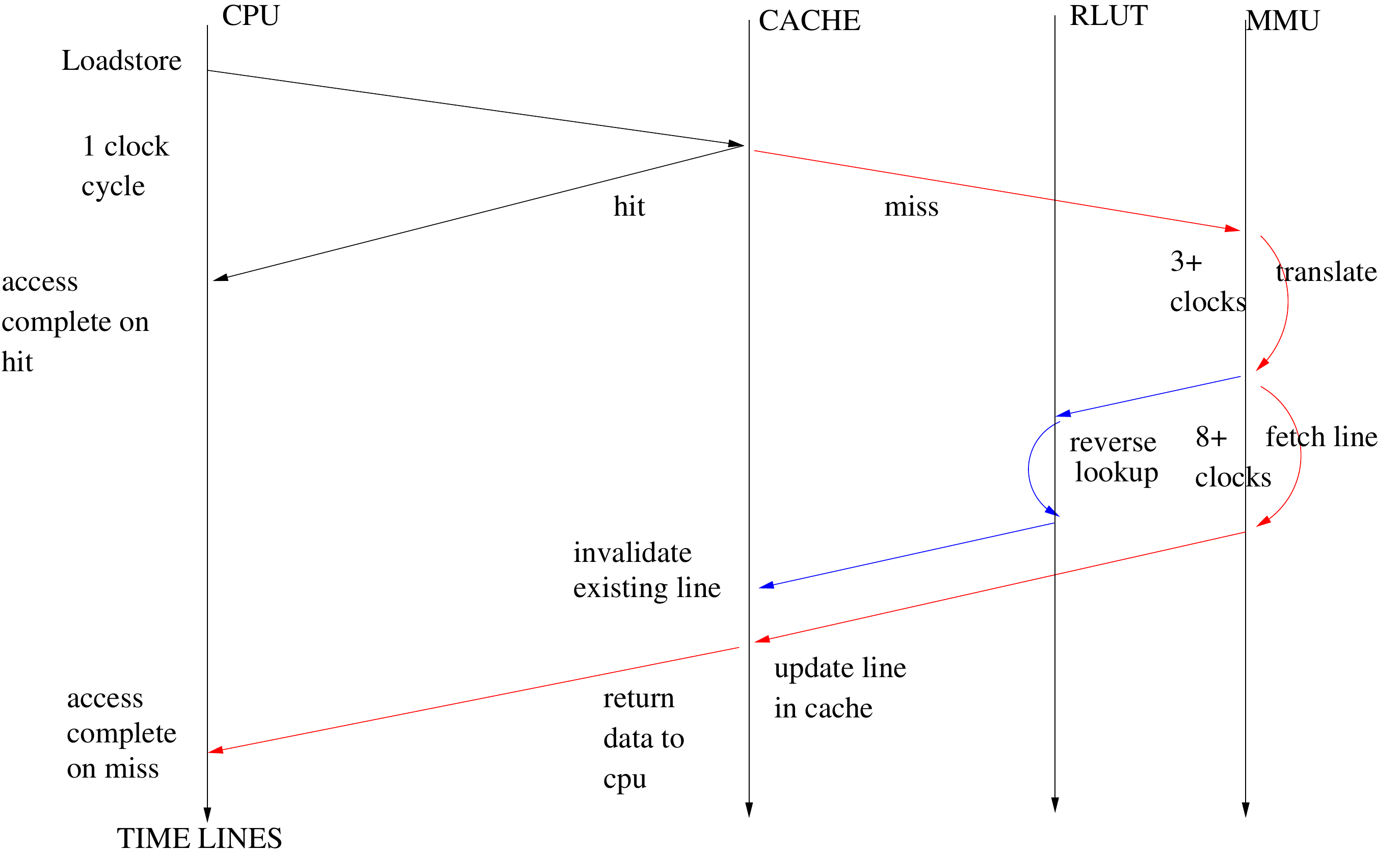}
  \caption{Activity in the memory access pipeline after introduction of the RLUT ($S=1$).}
  \label{fig:MemAccessFlowRlut}
\end{figure}

Thus, it is possible to maintain the invariant described above 
by doing a reverse lookup on the 
just translated physical address in the MMU and sending an invalidation
message to the cache asking it to invalidate the reverse translated 
virtual address, so that the invariant is maintained when 
the new virtual address is added to the cache after the line is fetched.
This is illustrated in Figure \ref{fig:MemAccessFlowRlut} for the case $S=1$.   
A reverse lookup table (RLUT) is introduced between the MMU and the
cache.  This reverse lookup table is organized as an 8-way associative
memory (for a 32 kilo-byte VIVT-DM cache, as discussed above).   Note that there is no lengthening
of the critical path in the memory access, because the application of
the RLUT driven action is happening concurrently with the fetch of
the cache line from physical memory.  In fact, we observe that the
time budget available for the reverse table lookup is 8 clock cycles,
which is substantial.

\section{Incorporating cache coherency related invalidations into the scheme}

\begin{figure}
  \centering
  \includegraphics[width=12cm]{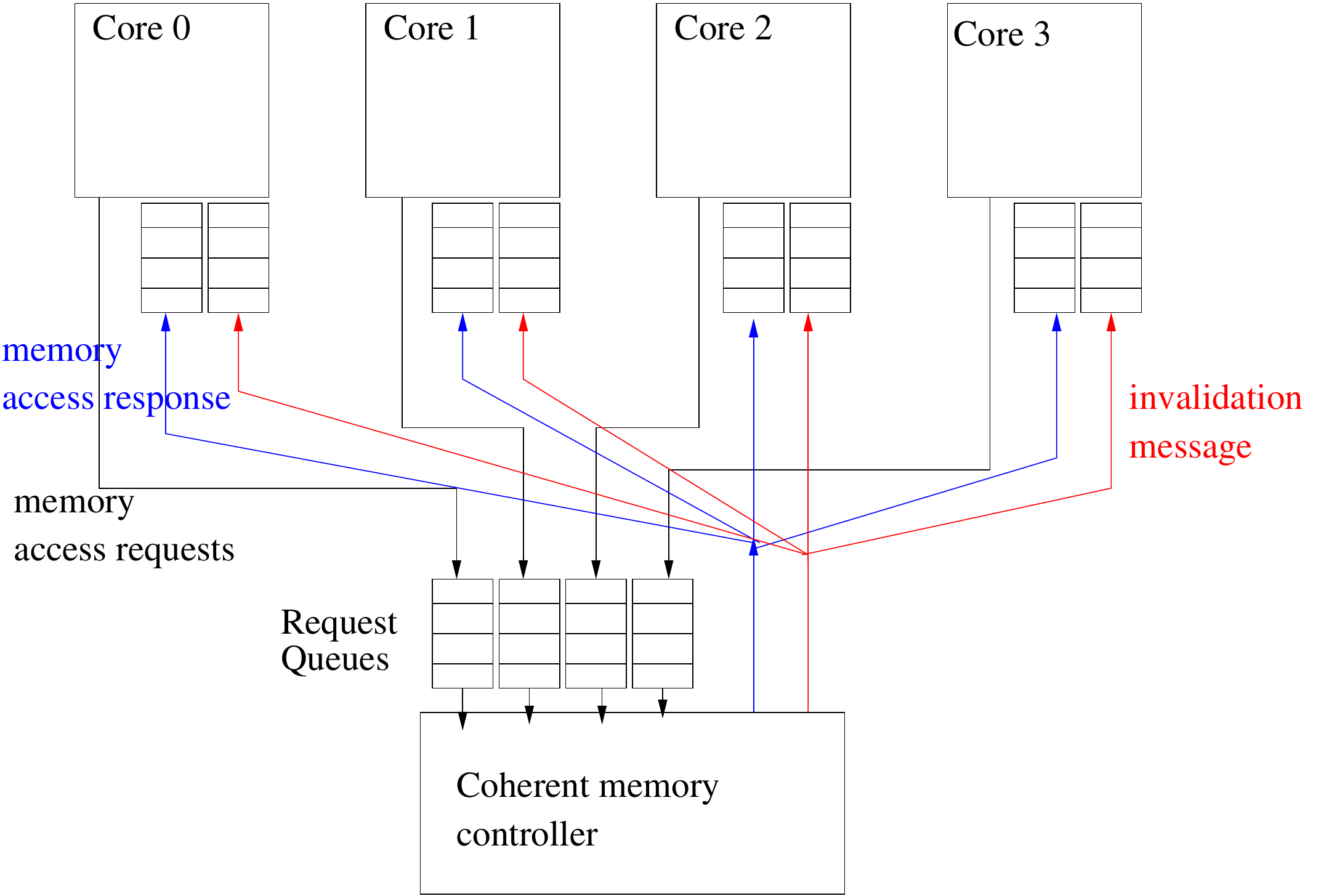}
  \caption{Four core AJIT system.}
  \label{fig:FourCoreSystem}
\end{figure}

In the AJIT-32 multi-core environment, invalidation messages are generated by 
a cache-coherent memory controller.   This is illustrated in Figure \ref{fig:FourCoreSystem}
for a four core system.

\begin{figure}
  \centering
  \includegraphics[width=12cm]{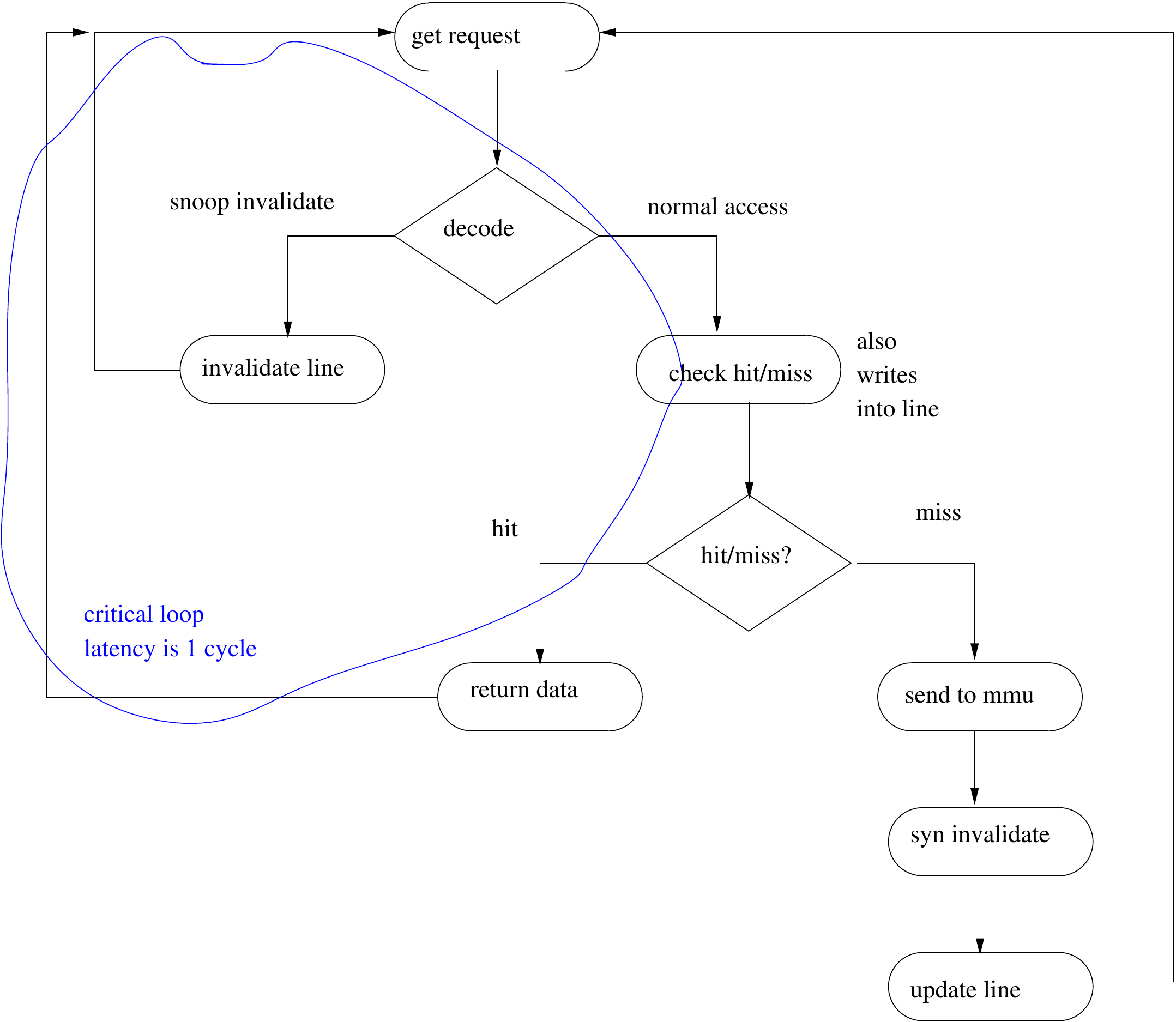}
  \caption{Cache controller state flow-chart}
  \label{fig:CacheControllerFlowchart}
\end{figure}

As far as a single core is concerned, we observe that it receives invalidation
messages which are not synchronous with its internal memory accesses.   These
external invalidation messages are synchronized to the cache access cycles by
the cache controller.   The cache controller samples the invalidate message
only when it has no miss pending.   This ensures that there are no race conditions
in the cache line invalidation due to the invalidation message racing ahead of
an ongoing miss access \cite{ref:SnoopProtocols}.   A simplified description of the
state machine implemented in the cache controller is shown in Figure \ref{fig:CacheControllerFlowchart}.
The behaviour of the state machine can be understood as follows:
\begin{itemize}
\item The cache controller accepts a new request.
\item If it is a coherence invalidate request then it
is immediately applied.  Otherwise, if a normal memory access is requested,
the cache controller proceeds to execute the memory access.
\item If a normal memory access is started, the tags and data arrays are accessed
in parallel and the hit status is computed.  If there is a hit, the 
result is reported immediately.
\item If there is a miss, a line access request is sent to the MMU.  The controller
then waits for the synonym invalidate message from the MMU (as indicated in Figure \ref{fig:MemAccessFlowRlut}).
\item After the  arrival of the synonym invalidate mesage, the cache controller
waits for the cache line to be received, updates the cache and is then ready
for the next request.
\end{itemize}

The critical loop is the request-hit-request loop, which has a loop latency
of one clock cycle, so that on hits, the cache serves one request per clock
cycle.

\begin{figure}
  \centering
  \includegraphics[width=12cm]{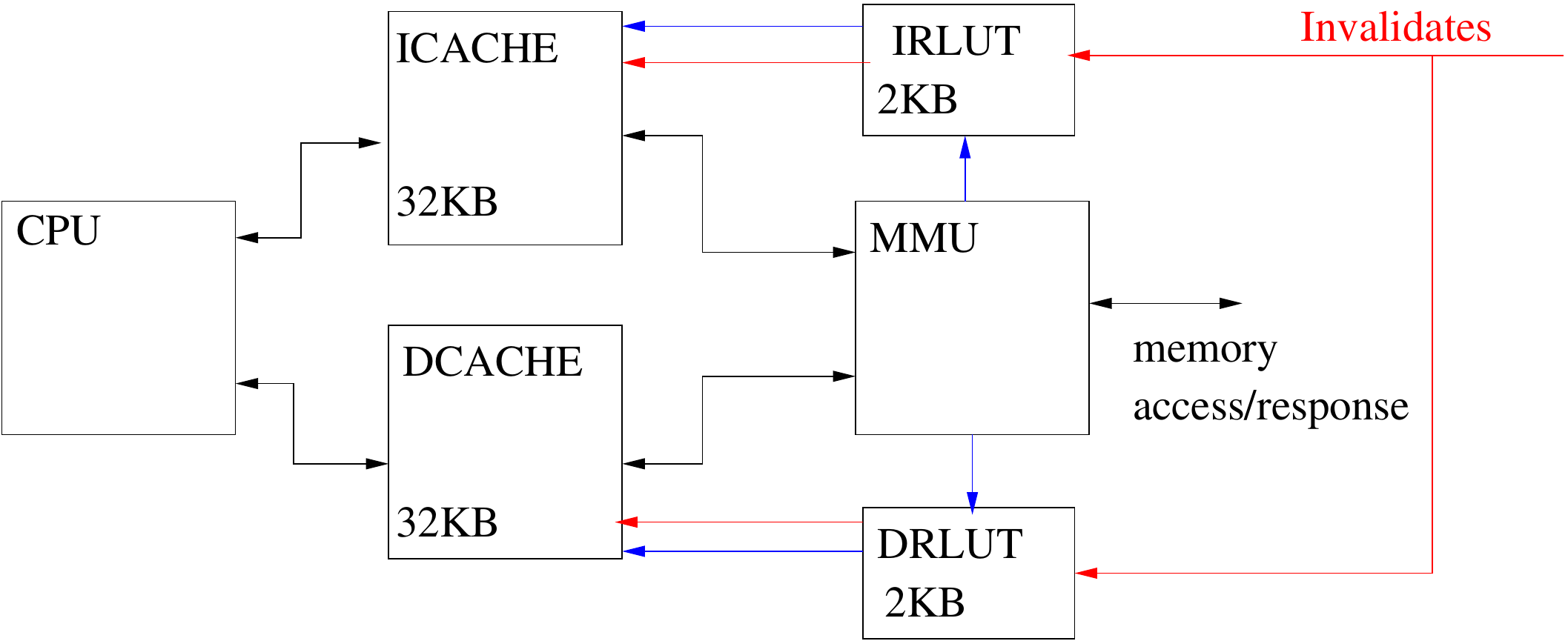}
  \caption{Final memory subsystem with RLUT.}
  \label{fig:FinalMemSubsystem}
\end{figure}

The final architecture of the memory subsystem in the AJIT core is illustrated
in Figure \ref{fig:FinalMemSubsystem}.   

\section{Hardware implementation characteristics of the RLUT for a $1$ synonym safe VIVT-DM cache}

\begin{figure}
  \centering
  \includegraphics[width=12cm]{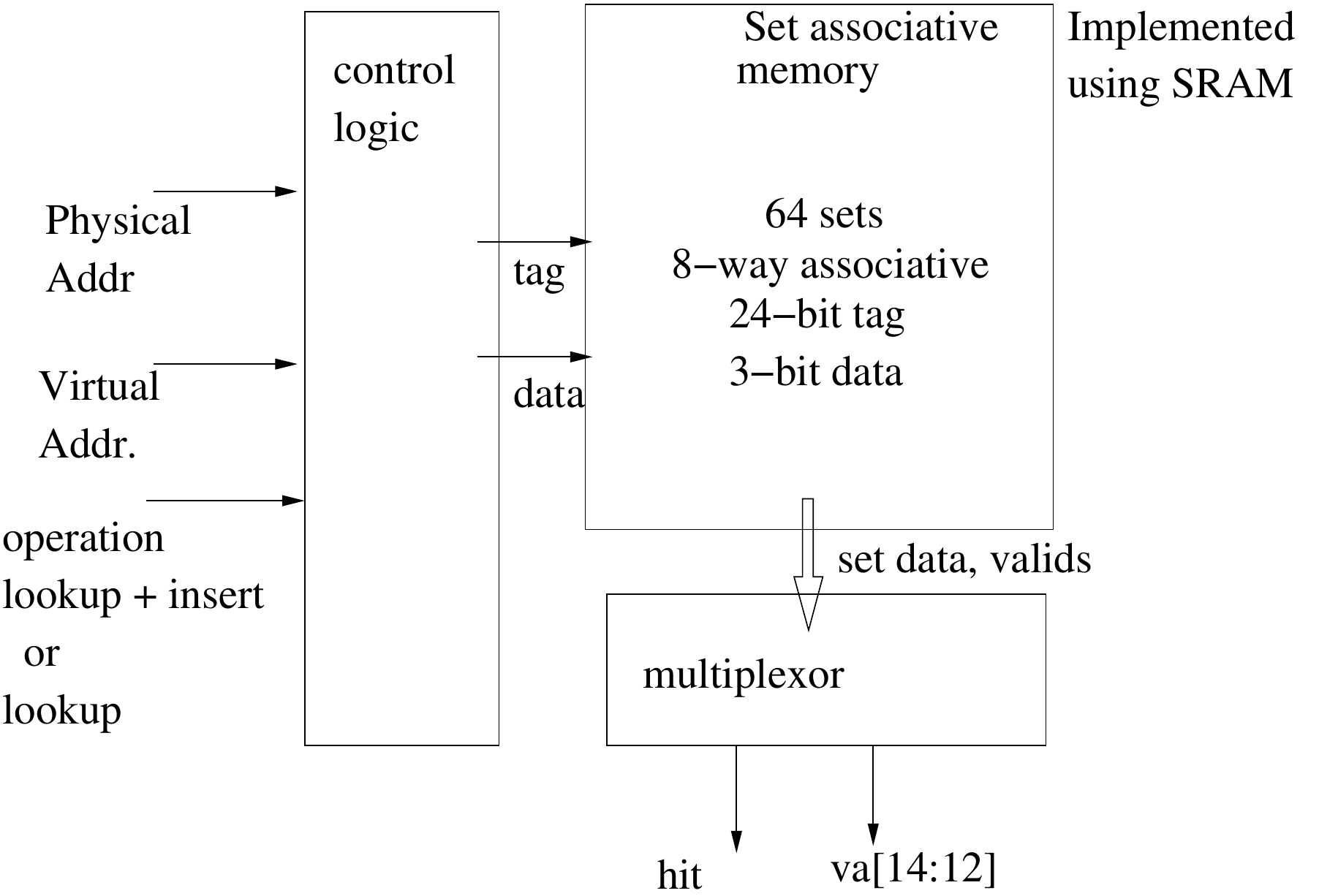}
  \caption{RLUT architecture}
  \label{fig:RlutArch}
\end{figure}

In Figure \ref{fig:RlutArch}, we show the architecture of the RLUT used in the
AJIT processor core (32 kilo-byte caches, 64-byte line, 1 synonym safe VIVT-DM).   
The RLUT supports the following operations.
\begin{itemize}
\item lookup and insert:  a (physical,virtual) address pair  $(P,V)$ is presented to the RLUT.  The
RLUT does a lookup in the set associative memory using $P$ as the tag and 
follows this up with an insert of $V$ as data for
the physical address $P$.  This operation takes two clock cycles and is not pipelined.
Note that this operation is applied only on a cache miss, and is a relatively rare event.
\item lookup: a physical address $P$ is presented to the RLUT. The RLUT does a lookup in the
set associative memory using $P$ and determines if there is a legal virtual address $W$ in 
the RLUT which is mapped to $P$.  This operation is fully pipelined.  Note that this
operation is triggered by snoop invalidates, and is common.
\end{itemize}

This RLUT implementation has been verified as part of the AJIT processor core on
an FPGA prototyping board (VC709) at a clock frequency of 100MHz.

The RLUT's are implemented using  an 8-way set-associative
cache, with 64 sets.   The entire physical address in the AJIT processor is
36 bits wide.  Since the line width is 64 bytes, and since there are 64 sets,
the RLUT set index is calculated using bits [11:6] of the physical address, and
the RLUT tag consists of bits [35:12] of the physical address.   The reverse
translation data consists of the $S$ virtual address of the line to which a physical address
can be mapped.   Coming to the $S$ virtual line addresses which can map to a
physical address $P$, we observe that it is sufficient to store just the {\em index}
of the virtual address line which maps to $P$.  For the numbers in consideration
(32 kilo-byte cache with 64 byte lines), a virtual address $V$ can be stored in
the line with index $V[14:6]$.  Of these bits, we know that if virtual
address $V$ is mapped to $P$, then $V[11:6] = P[11:6]$.   Thus, it is enough
to just store bits of $V[14:12]$ of $V$ in the RLUT.
Thus, for an $S$ synonym safe 32 kilo-byte VIVT cache, we need to
store
\begin{displaymath}
512 \times (24 + (3 \times S))
\end{displaymath}
bits of information in each RLUT.

In Figure \ref{fig:BitCount}, we present the number of bytes required for 
each RLUT as the cache size is varied (assuming that the page size is 4 kilo-bytes).  Note
that for a page size of 4 kilo-bytes, we do not need an RLUT for cache sizes of 4 kilo-bytes or
smaller.
\begin{figure}
\begin{verbatim}
-----------------------------------------
  Cache-size      S        Bytes-needed
-----------------------------------------
   4KB            1           0 
   8KB            1           432 
  16KB            1           864
  32KB            1           1728
   4KB            2           0
   8KB            2           480
  16KB            2           960
  32KB            2           1920
-----------------------------------------
\end{verbatim}
\caption{Number of memory bytes needed in each RLUT as a function of cache size and $S$}
\label{fig:BitCount}
\end{figure}
The general formula for the number of bytes needed is 
\begin{displaymath}
((CacheSize/64) \times (24 + (3 \times S)))/8
\end{displaymath}
when the cache size is $> 4$ kilo-bytes.

\section{Conclusions}

In this paper, we have presented a simple scheme for implementing synonym free,
coherent VIVT caches.  For a  given $S$, we ensure that there are at most
$S$ synonyms for a physical address present in the VIVT cache.  Keeping
$S=1$ yields the highest performance in the cache and also simplifies invalidation
actions which may arise due to writes to synonym physical addresses or
which arise as a result of cache coherence snoop activity.    For an
implementation of the scheme with $S=1$ on a direct-mapped 32 kilo-byte 
VIVT cache, the memory overhead required is 5.3\% and there is no impact on
the memory access latencies.   Further, snoop invalidation lookups are
completed in 1 clock cycle with a throughput of 1 lookup/clock-cycle, while
RLUT updates occur only on cache misses, and are completed in two clock cycles, concurrent
with the fetching of the line from main memory.
Thus, the best properties of VIVT caches
can be  preserved while providing a synonym free, coherent multi-core
ready processor core with no limits on the size of the L1 cache.

\bibliography{refs}

\end{document}